# Using causal networks to represent the targets of resource coordination

Eric Kuo, Nolan K. Weinlader, Benjamin M. Rottman, and Timothy J. Nokes-Malach

*Learning Research & Development Center, University of Pittsburgh, 3939 O'Hara St., Pittsburgh, PA, 15224*

The resources framework emphasizes the potential productivity of students' intuitions for constructing a canonical understanding of physics. It models learning as the progressive coordination and refinement of these resources. Yet, there is a lack of theoretical clarity about how resources should be coordinated and refined to align with canonical physics. We present causal network diagrams as a tool for representing the targets of research coordination. As an example, we compare student reasoning about projectile motion to the causal network describing that motion. We argue that the causal networks make manifest and explicit two types of resource coordination required to achieve a correct physical understanding: (i) integrating additional causal influences and mediators and (ii) using qualitative logic to draw valid inferences.

## I. INTRODUCTION

One key way that a theory of cognitive resources [1] (and other knowledge-in-pieces descriptions of learning [2]) differs from misconceptions-based interpretations of student thinking is in interpreting incorrect explanations. A misconceptions view treats incorrect explanations as errors that must be eliminated and replaced with correct conceptual knowledge [3]. On the other hand, a resources framework suggests that students possess ideas continuous with correct physics, even if they answer a question incorrectly. Accordingly, learning involves improved coordination of these existing resources. Yet, there is little theoretical clarity about how a resources framework theoretically describes and instructionally prescribes this improved coordination.

In this paper, we present *causal networks* (also called Bayes nets) [4] as a graphical formalism for representing physical models and demonstrate how this formalism makes explicit and manifest aspects of resource coordination necessary for aligning student ideas with canonical physics. We argue that formalizing the causal physics model affords explicit description of how student resources can fit with the canonical physics to be learned.

Although this is not the first effort to graphically represent conceptual reasoning networks, our approach is novel in that it uses causal networks to represent the endpoints of learning: the canonical physics models to be learned. In contrast, previous efforts have sought to graphically represent student reasoning [5]. As such, our contribution is to provide a representation that can systematically describe the goals of physics learning and illustrate how even elements of incorrect explanations can form the basis of a correct physical understanding.

## II. CAUSAL NETWORKS

Causal networks represent cause-effect relationships through a network of nodes and links. They have been developed for making correct causal determinations in complex situations, especially ones where experimental manipulation is impossible. With respect to cognition, causal networks have been used to characterize how human reasoning does and does not align with normative causal models [6].

Here, we employ causal networks to represent canonical physical models. Figure 1 shows a causal network representing projectile motion. The nodes represent different physical properties (in this case, all physical quantities). The directional links in the network indicate the causal influences between variables.

While the kinematic equations specify the quantitative relations between variables, one benefit of this causal network is the formalization of *qualitative*, causal relations between variables. For example, the causal network in Fig. 1 indicates that increasing the launch angle (and holding initial speed constant) will increase the vertical component of the initial velocity, which will increase both the peak height and time in the air. However, the effect on horizontal range is qualitatively ambiguous, since increasing the launch angle also decreases the horizontal component of the velocity. These qualitative, causal relations are explicitly represented by the causal network formalism, so constructing this explicit representation of projectile motion provides a complete model for evaluating how student reasoning does or does not align with canonical physics.

There are two reasons to believe that causal networks should be a useful formalism for understanding physics learning. The first reason is that physical science can be understood to be the search for and continual development of causal explanations. Therefore, causal networks represent a core element of scientific models. The other reason is that many of students' intuitions, their resources for learning physics, exhibit causal character. Phenomenological primitives (p-prims) [7] lie at the base of an intuitive sense of causal mechanism. At their core, p-prims such as *force as a mover* and *Ohm's p-prim* (increased effort leads to more result & increased resistance leads to less result) are causal notions. Furthermore, diSessa and Sherin [2] theorized that people possess "causal nets" for drawing inferences from given information. They argued that much of the challenge of learning physics is in reconfiguring the causal net to align with canonical physics concepts. However, they did not formalize the structure of these causal nets and comparison between student reasoning and canonical physics was done in an ad-hoc, rather than systematic, manner.

Here, we build on their theory of causal nets by using the causal network formalism to represent canonical inference-making structures. We will compare student reasoning to these canonical physical models to highlight two dimensions of resource coordination for learning physics.

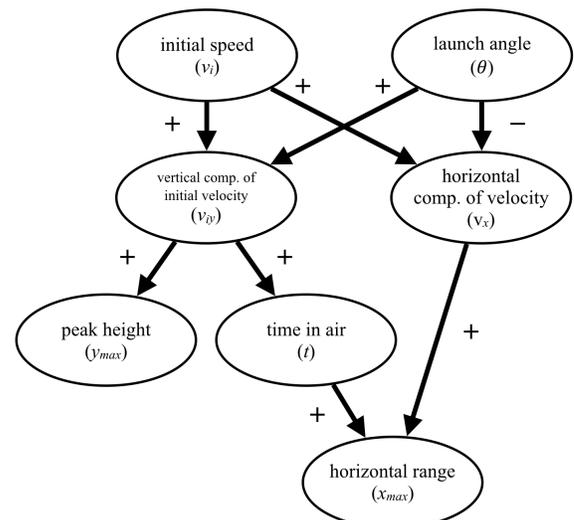

FIG 1. Causal network for projectile motion (constant $g$).

## III. CAUSAL NETWORKS SPECIFY DIMENSIONS OF RESOURCE COORDINATION – THE EXAMPLE OF PROJECTILE MOTION

To illustrate how causal networks can represent physics learning goals, we compare student reasoning on the following two questions (Fig. 2, taken from *smartPhysics* [8]) to the causal network of projectile motion. Correct explanations for these two questions can be illustrated by examining how qualitative differences between shells 1 and 2 propagate through the causal network in Figure 1. In Q1, shell 2 reaches a lower peak height. From this, we can infer that shell 2 must have had a smaller vertical component of its initial velocity, which therefore led to less time in the air. For Q2, the same argument pathway can be used to infer that the targets are hit at the same time. Using peak height to infer time in air is a clear projectile motion learning goal.

We use student responses to Two Boats Q1 and Q2 to explore the usefulness of causal networks for describing dimensions of resource coordination for learning physics. Sixteen undergraduate students who were enrolled in or had taken college physics were asked to come up with an explanation for why someone might believe each of the following answers was correct: target 1 gets hit first, target 2 gets hit first, and they are hit at the same time. From this prompt, students' responses may not reflect their own conceptual views. However, we take their explanations as plausible examples of how students can coordinate their resources for answering physics questions and explore the implications of these explanations for physics instruction.

### A. Coordination as integrating additional causal paths and mediators into existing reasoning

Consistent with the misconceptions perspective, one could view incorrect explanations as needing to be overturned in favor of a correct solution, such as using peak height to determine relative times in the air. However, the causal network for projectile motion suggests an alternative learning goal consistent with the resources perspective: continual integration of student resources into the canonical causal network.

For instance, on Q1, to provide an explanation for why target 2 would get hit first, 7 students reasoned that the launch angle for shell 2 was lower and/or the path of shell 2 was more direct. This *lower launch angle is less time in the air* resource is not correct generally, since the initial speeds could also differ, as in Q2. However, it is consistent with a valid causal pathway in the causal network for projectile motion (launch angle $\rightarrow$ vertical component of initial velocity $\rightarrow$ time in air). Here, in the special case of equal initial speeds (Q1), this path is the only one that influences the time in the air, so this explanation leads to a correct conclusion.

If the goal is to help students answer such questions reliably, an instructor may wish to promote the use of peak height to determine the time in the air over this approach. However, if the goal of learning physics is to help students construct complete physical models of phenomena, there are two ways in which we might want students to elaborate on *lower launch angle is less time in the air*.

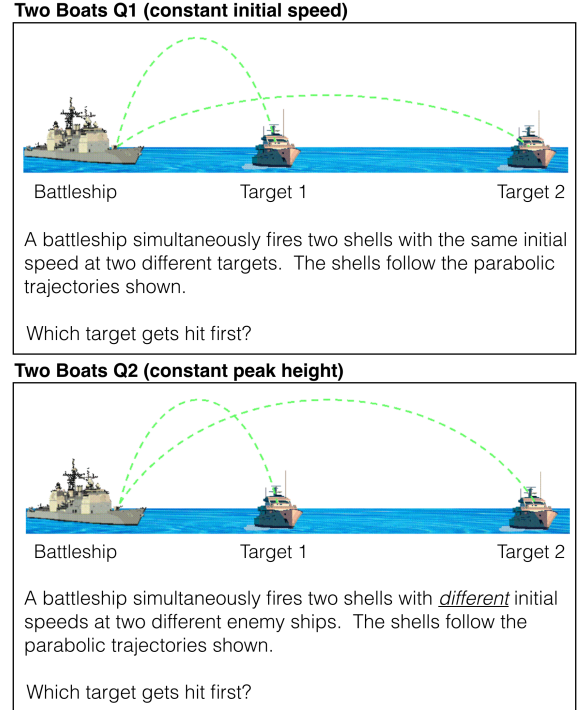

FIG 2. Two qualitative problems asking students to determine the relative flight times of two projectiles.

The first elaboration is to explicitly recognize and consider multiple causal influences. A complete understanding of projectile motion would recognize that both launch angle *and* initial speed have a causal influence on time in the air. The influence of initial speed is explicitly represented in the causal network by this pathway: initial speed $\rightarrow$ vertical component of initial velocity $\rightarrow$ time in air. Recognizing the causal effect of initial speed on time in the air is especially critical for transitioning from Q1 to Q2, since the validity of the *lower angle is less time in the air* reasoning on Q1 is conditional on equal initial speeds for the two shells, a condition that doesn't hold for Q2.

Second, we ideally want students to be able to specify the mediators of these causal paths. Here, the mediator of initial speed and launch angle on time in the air is $v_{iy}$. Specifying this mediator is not only a more complete description of the physics, but also allows for valid inference in situations where considering initial speed and launch angle cannot determine time in the air (such as in Q2, where shell 1 has a higher launch angle and lower initial speed).

Existing analytic methods using a resources perspective may have identified *lower angle is less time in the air* as a potentially productive resource in learning physics, but there has not existed a formal framework for describing how this idea could be systematically integrated into a canonical understanding of projectile motion. The causal network formalism represents the role of this explanation in the canonical physical model explicitly and illustrates how integration with additional causal influences and detailing mediators of this causal path can produce a more complete causal understanding.

Even when students' reasoning leads to incorrect conclusions, this reasoning may reveal resources that align with the canonical causal network. Considering responses for Q1 and Q2, all 16 students interviewed gave an explanation that, because target 1 was closer to the battleship (or shell 1 had a smaller horizontal range), it would get hit first. This *more distance is more time* resource has been seen in other kinematics problem contexts [9]. For projectile motion, the causal network illustrates why more horizontal range does not necessarily imply more time: more horizontal range could be due to more time in the air and/or a greater $v_x$ (Fig. 1). Similarly omitting one causal factor, for Q2, 2 students explained that target 2 would be hit first because shell 2 has a greater horizontal speed, failing to consider the difference in horizontal distances traveled by the two shells.

Again, since the *more distance is more time* explanation is invalid here, one instructional approach would be to confront this incorrect approach and dissuade students from using it, bringing students' attention instead to peak height and $v_{iy}$ as valid indicators of the time in the air. On the other hand, the causal network indicates that *more distance is more time* represents a valid physical relationship, as does *more speed is more distance*. In terms of learning the complete physical model, the implication is that students should learn to integrate both of these causal relations into an understanding of projectile motion. Recognizing and integrating resources that align with parts of the causal network can focus instruction on the aim of helping students construct a more complete, coherent model of the physical system, rather than one that is merely sufficient to answer the question at hand.

### B. Coordination as drawing valid qualitative inferences from the causal network

From a resources perspective, much of the attention has been on identification of sets of resources that students activate to construct their reasoning. Comparatively less attention has been paid toward understanding the reasoning required to make inferences when considering multiple causes and/or effects. The causal networks make the complexity of this reasoning more explicit.

For example, take the previously discussed relations between time, horizontal speed, and horizontal distance. For Q1, shell 2 has a greater horizontal speed and travels a greater horizontal distance than shell 1. What can be said about the time in the air of shell 2 compared to shell 1?

TABLE I. Possible inferences on changes in time given increases (+)/decreases (−) in horizontal speed and range.

|  |  | Change in horizontal range | | |
|---|---|---|---|---|
|  |  | + | 0 | − |
| Change in horizontal speed | + | ? | − | − |
|  | 0 | + | 0 | − |
|  | − | + | + | ? |

Table 1 shows the qualitative logic for inferring changes in time (a cause) from changes in horizontal velocity (a cause) and changes in horizontal range (an effect). Definite judgments about time can be made when the change in horizontal speed does not qualitatively match the change in horizontal range. For example, when horizontal speed increases (+) and horizontal range stays constant (0) or decreases (−), then we can infer that time decreases (−). However, there are also situations where no inference can be made. For example, 6 students used a flawed compensation argument on Q2: projectile 2 travels a greater (horizontal) range and also has a faster (horizontal) speed, so they concluded the targets will be hit at the same time. In this case, students are coordinating information about distance and speed to draw inferences about time, but the inferential logic is flawed. Table 1 indicates that when shell 2 has a greater speed (+) and greater range (+), no qualitative conclusion can be made about the time (?). In this way, causal networks highlight the importance of valid qualitative logic, even when all relevant causal relations are considered.

### IV. DISCUSSION

We present causal networks as a representation of canonical physics that indicates how even students' incorrect explanations contain elements of correct physical understanding. We see causal networks clarifying several dimensions of physics learning suggested by a resources framework.

First, causal networks can bring clarity to a common mantra of the resources framework: "Resources are neither correct nor incorrect. Learning physics means learning to apply these resources in canonically correct ways." If the focus is on teaching students the correct solutions to particular physics questions, incorrect explanations (such as "target 1 is hit first, because it's closer" on the Two Boats questions) can be dismissed as invalid. However, if the goal is to learn and reason with the complete physical models described by these causal networks, learning physics is more complex: it requires integrating the valid relations that these incorrect explanations reflect into one's causal reasoning. This is consistent with instruction focused on *knowledge refinement* [10], which explicitly aims to integrate incorrectly applied intuitions into a correct physical understanding. This approach also has a conceptual benefit of helping students construct a coherent knowledge structure and an epistemological benefit of helping students view their intuitions as productive, rather than harmful, for learning physics.

Second, causal networks not only highlight how qualitative resources can be productive for correct physical understanding, but they also illustrate that drawing inferences from multiple qualitative influences is a non-trivial reasoning task. Explanations such as speed/distance compensation illustrate that even when the correct conceptual elements are in play, correct inference may depend on valid qualitative logic. Uncovering student resources for qualitative logic and understanding how students can use these resources for drawing inferences is an important topic for future research.

Third, causal networks can indicate the importance of differentiating one's causal knowledge in learning physics. Although distance, time, and speed are universally related, the causal network for projectile motion explicitly represents how the *causal* relations between these kinematic quantities differ for horizontal and vertical components of the motion. Horizontally, time and speed have causal influence on distance, but vertically the initial speed is the common cause of distance and time. Therefore, learning to differentiate causal relations for horizontal and vertical motion may be a key step in learning about projectile motion. Relatedly, this explicit attention on causality highlights a difficulty in learning physics from the equations alone: the same equation can represent different causal situations. That is, the independent and dependent variables can change across the problem contexts. Consider the equation $x = x_0 + v_0 t + \frac{1}{2} a t^2$. In the case of tossing a ball upward to its peak, $v_0$ and $x_0$ are the independent variables, and $x$ and $t$ are the dependent outcomes ($a$ is held constant). On the other hand, in describing a drag race, $x_0$, $v_0$, $a$, and $x$ are all independent variables (set by the rules of the race and the cars' construction), and $t$ is the dependent outcome. The causal relations between quantities are a key piece of the physical understanding that students need to integrate with the mathematical constraints for correct physics problem solving [11].

For these reasons, we believe that the causal network formalism presents a path forward for specifying the productive elements of incorrect explanations. This work suggests new directions for instruction while also improving the clarity of the resources framework and its implications.

---

[1] D. Hammer, A. Elby, R. E. Scherr, and E. F. Redish, in *Transfer of Learning from a Modern Multidisciplinary Perspective*, edited by J. Mestre (Information Age Publishing, Greenwich, CT, 2005), pp. 89–120.
[2] A. A. diSessa and B. L. Sherin, Int. J. Sci. Ed. **20**, 1155 (1998).
[3] J. P. Smith, A. A. diSessa, and J. Roschelle, J. Learn. Sci. **3**, 115 (1993).
[4] J. Pearl, *Causality: Models, Reasoning and Inference*, 2nd edition (Cambridge University Press, Cambridge, U.K. ; New York, 2009).
[5] M. C. Wittmann, Phys. Rev. ST Phys. Educ. Res. **2**, 020105 (2006).
[6] B. M. Rottman and R. Hastie, Psychol. Bull. **140**, 109 (2014).
[7] A. A. diSessa, Cogn. Inst. **10**, 105 (1993).
[8] G. Gladding, M. Selen, and T. Stelzer, *SmartPhysics* (MacMillan Education, 2015).
[9] B. W. Frank, S. E. Kanim, and L. S. Gomez, Phys. Rev. ST Phys. Educ. Res. **4**, 020102 (2008).
[10] D. Hammer and A. Elby, J. Learn. Sci. **12**, 53 (2003).
[11] E. F. Redish and E. Kuo, Sci. Educ. **24**, 561 (2015).